\documentclass[prl,aps,twocolumn,superscriptaddress]{revtex4-1}

\usepackage{graphicx}
\usepackage{amssymb,amsfonts,amsmath}
\usepackage{color}
\usepackage{ulem}
\usepackage[hidelinks]{hyperref}
\usepackage{bbm}
\usepackage{bbold}

\graphicspath{{./figures/}}

\usepackage{tikz}
\usetikzlibrary{shapes}
\usetikzlibrary{patterns}
\usetikzlibrary{angles, quotes}

\newcommand{\rv}{{\mathbf r}}

\newcommand{\ev}{{\bf e}}

\newcommand{\e}{{\rm e}}

\newcommand{\Jv}{{\bf J}}

\newcommand{\Fv}{{\bf F}}
\newcommand{\Gv}{{\bf G}}
\newcommand{\fv}{{\bf f}}

\newcommand{\vel}{{\bf v}}

\newcommand{\msphantom}[1]{$\ldots$}

\newcommand{\eqr}[1]{Eq.~\eqref{#1}}

\newcommand{\mydelete}[1]{{}}

\newcommand{\rmint}{{\rm int}}

\newcommand{\rmext}{{\rm ext}}

\newcommand{\sgn}{\operatorname{sgn}}
\newcommand{\alphasup}{{(\alpha)}}
\newcommand{\calG}{{\cal G}}

\begin{document}

\title{Nonequilibrium scaling of drag forces in counterdriven fluid mixtures}

\author{Jonas K\"oglmayr}
\affiliation{Theoretische Physik II, Physikalisches Institut, 
  Universit{\"a}t Bayreuth, D-95447 Bayreuth, Germany}

\author{Florian Samm\"uller}
\affiliation{Theoretische Physik II, Physikalisches Institut, 
  Universit{\"a}t Bayreuth, D-95447 Bayreuth, Germany}

\author{Matthias Schmidt}
\email{Matthias.Schmidt@uni-bayreuth.de}
\affiliation{Theoretische Physik II, Physikalisches Institut, 
  Universit{\"a}t Bayreuth, D-95447 Bayreuth, Germany}

\date{29 May 2026, revision: 6 August 2026}

\begin{abstract}
  We address the effective nonequilibrium drag force field that
  emerges from the microscopic interparticle interactions in steady
  states of counterdriven binary fluid mixtures. Using power
  functional scaling arguments for adaptive Brownian dynamics computer
  simulation results, we establish quantitatively the crossover
  between near-equilibrium linear response and far-nonequilibrium
  square root asymptotics. An algebraic expression captures both
  limiting cases and remains applicable in the crossover regime. Using
  simulation results as benchmarks, we verify that a local power
  functional approximation based on the scaling law reproduces the
  spatial nonequilibrium structure formation in inhomogenously driven
  systems. The crossover scenario transcends dynamical density
  functional theory and it sheds light on general nonequilibrium
  scaling of driven fluids.
\end{abstract}

\maketitle

The physics of fluid mixtures can be considerably richer than the
phenomenology that pure, one-component systems display
\cite{hansen2013}. Recently studied collective phenomena that are
specific to multi-component systems include composition-dependent
phase instabilities \cite{thewes2023, thewes2024} and mobility effects
in nonequilibrium~\cite{akaberian2023}. When the two particle species
in a {\it binary} mixture are driven in opposite directions, then a
range of nonequilibrium structure formation effects occur, including
laning and jamming~\cite{dzubiella2002laning,
  chakrabarti2003laning, chakrabarti2004laning, yu2024laning,
  geigenfeind2020laning}.
To find fluid-fluid phase separation in equilibrium requires typically
the presence of interparticle attraction, with the physics of
liquid-liquid and liquid-gas phase transitions and their associated
interfacial wetting phenomena being systematically accessible via
equilibrium density functional theory \cite{telodagama1983one,
  telodagama1983two, tarazona1985interface, hadjiagapiou1985}.
Binary Lennard-Jones models \cite{wilding1997, schmid2001wetting,
  wilding2002} are prototypical \cite{telodagama1983one,
  telodagama1983two, tarazona1985interface, hadjiagapiou1985,
  wilding1997, schmid2001wetting, wilding2002} and these were used
recently to demonstrate machine learning of accurate equilibrium
density functionals that extrapolate significantly beyond the physics
encountered during the simulation-based
training~\cite{robitschko2025mixShort, zhou2026azeoptropic}, as also
demonstrated for electrolytes \cite{bui2024neuralrpm,
  bui2025dielectrocapillarity}.

The effective forces that generate nonequilibrium laning were
analyzed~\cite{geigenfeind2020laning} according to power functional
theory \cite{schmidt2022rmp}. This universal approach for general
nonequilibrium dynamics is based on a formally exact one-body
minimization principle. Workable approximations for the central
functional dependencies were constructed analytically to describe the
dependence on the velocity gradient of the occurring flow patterns
\cite{stuhlmueller2018prl, delasheras2018velocityGradient,
  delasheras2020fourForces}, nonlocal memory convolution kernels
\cite{treffenstaedt2020shear, treffenstaedt2021dtpl,
  treffenstaedt2022dtpl}, and the behaviour under shear
\cite{jahreis2019shear}.  The recent machine learning of neural
nonequilibrium force functionals \cite{delasheras2023perspective,
  zimmermann2024ml} allows one in particular to go beyond the
limitations of dynamical density functional theory \cite{evans1979,
  tevrugt2020review}, which is a currently much used approach, see
e.g.~Refs.~\cite{millswillaims2024, tschopp2024ddft2, ram2025ddft}.

Here we aim to make progress in the description of nonequilibrium
fluids by investigating the specific type of flows that occur in
counterdriven binary mixtures. We consider both spatially homogeneous
and inhomogeneous systems over a broad range of dilute and dense
conditions under weak and strong driving. While we scan a wide set of
parameter choices, we avoid deliberately those regimes where laning
and jamming effects occur \cite{dzubiella2002laning,
  chakrabarti2003laning, chakrabarti2004laning, yu2024laning,
  geigenfeind2020laning}, and aim for a full quantitative description
of the drag force field that emerges from the interparticle collisions
between two counterflowing fluid species.  We identify a
well-characterized crossover regime that connects near-equilibrium
linear response with far-nonequilibrium scaling that we identify under
strong driving. The scaling of the drag force as a function of the
counterflow velocity is linear near equilibrium and it has square-root
dependence far in nonequilibrium.
The development of the nonequilibrium scaling theory is based solely
on analyzing systems that are spatially homogeneous, such that the two
partial (i.e.~species-resolved) density profiles are spatially
constant. In such situations, dynamical density functional theory
predicts per construction vanishing mean interparticle forces and
hence it will not account for the observations.  In contrast power
functional scaling gives accurate quantitative account of the drag
forces and we demonstrate that the approach transcends the homogeneous
driving conditions and describes quantitatively spatially
inhomogenenous nonequilibrium.

We use overdamped Brownian dynamics \cite{hansen2013,dhont1996book} to
model the microscopic many-body dynamics of the counterflowing mixture
and implement the trajectory-based stochastic Langevin equations via
adaptive Brownian dynamics simulations \cite{sammueller2021}.  For
overdamped dynamics the only relevant microscopic degrees of freedom
are the positions $\rv_1,\ldots,\rv_N\equiv \rv^N$ of the $N$
particles, with $\rv^N$ being a shorthand notation for all particle
coordinates, and we omit hydrodynamic interactions.  To discriminate between the different species we use
index sets ${\cal N}_\alpha$ that contain all particle indices $i$
that belong to species~$\alpha=1, 2$ in the binary mixtures.  The
system is driven out of equilibrium by species-resolved external force
fields $\fv_\rmext^\alphasup(\rv)$ that couple individually to each
species~$\alpha$.  We are interested in steady states, where all mean
one-body quantities depend on position $\rv$ and are independent of
time $t$. The reduction from three-dimensional to planar geometry is
specified later.

The interparticle repulsion is modelled by the Weeks-Chandler-Andersen
pair potential, which is a truncated and shifted, purely repulsive
Lennard-Jones potential given by $\phi(r)=4 \epsilon
[(\sigma/r)^{12}-(\sigma/r)^6]+\epsilon$ for $r<2^{1/6}\sigma$ and
zero otherwise; $\epsilon$ is the energy scale and the
lengthscale~$\sigma$ sets the particle size. The particles of the two
different species share the same interaction potential, hence we
consider a `painted particle' model \cite{thewes2023,
  geigenfeind2020laning}, where the interparticle potential is
$u(\rv^N)=\sum_{i,j(\neq i)}\phi(|\rv_i-\rv_j|)/2$.
The individual species are solely discriminated by the strength and
the spatial form of the external force fields
$\fv_\rmext^\alphasup(\rv)$ that act on the different species
$\alpha=1,2$.
We  use a three-dimensional periodic cubic simulation box with linear size
$L=10\sigma$ and consider partial particle numbers in the range
$N_\alpha=30-300$ for each species $\alpha=1,2$.  The temperature $T$
is set such that $k_BT/\epsilon=0.5$, where $k_B$ denotes Boltzmann's
constant.  For each pair of partial density values considered, we
realize 50 different driving strengths. For each system, after an
initial relaxation period, steady state properties are sampled for a
duration of $1000\tau$, where $\tau=\sigma^2\gamma/\epsilon$ is the
microscopic time scale, with $\gamma$ the common friction constant and
$D=k_BT/\gamma$ the associated free diffusion constant.
The numerical time evolution of the overdamped Langevin equations is implemented via adaptive Brownian dynamics, which facilitates automatic adjustment of the time discretization interval and hence improves upon accuracy and stability limitations of the common Euler-Maruyama scheme. This is particularly relevant in light of the wide range of conditions investigated and for controlling the numerical error of interparticle forces that we aim to analyze. A detailed description of the method is given in Ref.~\cite{sammueller2021}.

\begin{figure}[!t]
  \vspace{1mm}
  \includegraphics{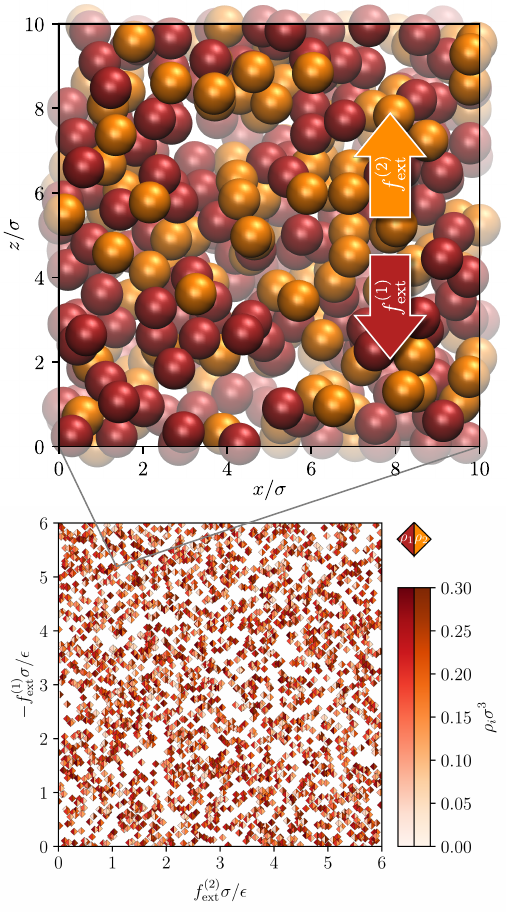}
  \caption{Top panel: Representative simulation snapshot of the binary
    mixture under counterdrive. The particles of species~1 (dark red
    spheres) and of species~2 (light orange spheres) are driven in
    opposite directions along the $z$-axis by species-dependent
    external force fields $f^{(1)}_\rmext(z)$ and $f^{(2)}_\rmext(z)$,
    which are first taken to be spatially uniform and are later
    considered to be position-dependent on the $z$-coordinate.  The
    system is on average homogeneous in the lateral $x$- and
    $y$-directions and lengths are scaled with the common particle
    size $\sigma$. Bottom panel: Illustration of the range of
    simulated systems with spatially homogeneous driving, shown across
    the plane of scaled force constants $f^{(1)}_\rmext
    \sigma/\epsilon$ and $f^{(2)}_\rmext \sigma/\epsilon$ with
    colour-coded partial densities $\rho_1 \sigma^3$ and $\rho_2
    \sigma^3$ (semi-diamond symbols).}
\label{FIGsnapshot}
\end{figure}

We base the statistical mechanical description of the system on the
dynamical one-body picture \cite{hansen2013, schmidt2022rmp}. The
primary observables of interest are the partial density profile
$\rho_\alpha(\rv)$ and the partial interparticle force density
$\Fv_\rmint^\alphasup(\rv)$. These are respectively defined as
\cite{schmidt2022rmp}:
\begin{align}
  \rho_\alpha(\rv) &=
  \Big\langle
  \sum_{i\in{\cal N}_\alpha}\delta(\rv-\rv_i) \Big\rangle,
  \label{EQrhoAlphaDefinition}\\
   \Fv_\rmint^{(\alpha)}(\rv) &=
   -\Big \langle
   \sum_{i\in{\cal N}_\alpha}\delta(\rv-\rv_i)\nabla_i u(\rv^N)
   \Big\rangle,
   \label{EQFintAlphaDefinition}
\end{align}
where the angles denote the steady state average, the sums are over
(only) particles indices $i\in{\cal N}_\alpha$ to ensure the correct
attribution to species $\alpha=1,2$, the symbol $\delta(\,\cdot\,)$
indicates the Dirac distribution, and $\nabla_i$ is the dervative with
respect to particle position $\rv_i$. In general, imposing an external
force field generates mean motion, which is characterized by
nonvanishing one-body partial currents $\Jv_\alpha(\rv)$. For
overdamped Brownian systems the partial current is induced by
species-resolved diffusive, interparticle, and external contributions
according to the force density balance~\cite{schmidt2022rmp}:
\begin{align}
  \gamma \Jv_\alpha(\rv) &= -k_BT \nabla\rho_\alpha(\rv)
  +\Fv_\rmint^{(\alpha)}(\rv) + \rho_\alpha(\rv)
  \fv_\rmext^{(\alpha)}(\rv).
  \label{EQequationOfMotion}
\end{align}
For steady states the partial currents are divergence free, as follows
from the continuity equation
$\nabla\cdot\Jv_\alpha(\rv)=-\partial\rho_\alpha(\rv)/\partial t=0$.
The partial velocity profile of species $\alpha$ is the ratio
$\vel_\alpha(\rv)=\Jv_\alpha(\rv)/\rho_\alpha(\rv)$.
Power functional theory ascertains \cite{schmidt2022rmp} that
$\Fv_\rmint^\alphasup(\rv,[\rho_1,\rho_2,\vel_1,\vel_2])$ is a
universal functional of the partial density and velocity profiles, as
is indicated in the notation by the square brackets.  The dependence
on these fields is causal and the universality refers to the absence
of any functional dependence on the external force fields
$\fv_\rmext^{(1)}(\rv)$ and $\fv_\rmext^{(2)}(\rv)$.  That this type
of functional dependence holds is a nontrivial result that is not
apparent from \eqr{EQequationOfMotion} alone. We recall that power
functional theory is the extension of classical density functional
theory \cite{evans1979}, which establishes that in equilibrium, the
internal force profile is a universal functional of (only) the
densities, $\Fv^{(\alpha)}_\mathrm{int}(\rv, [\rho_1, \rho_2])$.

We first follow Geigenfeind {\it et al.}~\cite{geigenfeind2020laning}
to decompose the species-resolved interparticle force densities
\eqref{EQFintAlphaDefinition} into two separate contributions. Thereby
the species-independent interparticle force field $\fv_\rmint(\rv)$ is
defined as the contribution that acts identically on both species.
The counterforce (``differential force'' \cite{geigenfeind2020laning})
density $\Gv_\rmint(\rv)$ acts on the relative motion between the two
species. This decomposition constitutes a change of variables from
$\Fv_\rmint^{(1)}(\rv)$ and $\Fv_\rmint^{(2)}(\rv)$ to
$\fv_\rmint(\rv)$ and $\Gv_\rmint(\rv)$, which is defined by:
\begin{align}
  \Fv_\rmint^\alphasup(\rv) &= 
  \rho_\alpha(\rv) \fv_\rmint(\rv) \pm \Gv_\rmint(\rv),
  \quad \alpha=1,2,
  \label{EQFintalpha}
\end{align}
where $\pm$ refers to species $\alpha=1,2$. By summing over both
species in \eqr{EQFintalpha} one obtains by construction the total
interparticle force density $\Fv_\rmint(\rv)=\Fv_\rmint^{(1)}(\rv)
+\Fv_\rmint^{(2)}(\rv) =\rho(\rv)\fv_\rmint(\rv)$, with the total
density profile $\rho(\rv)=\rho_1(\rv)+\rho_2(\rv)$.  The counterforce
density $\Gv_\rmint(\rv)$ follows correspondingly by subtraction {(and
  multiplication by $1/2$) and thus~\cite{geigenfeind2020laning}:}
\begin{align}
  \fv_\rmint(\rv) &= \Fv_\rmint(\rv)/\rho(\rv),
  \label{EQfintGeneralDefinition}\\
  \Gv_\rmint(\rv) &= \big[
  \rho_2(\rv) \Fv^{(1)}_\rmint(\rv)
    -\rho_1(\rv) \Fv^{(2)}_\rmint(\rv)\big]/\rho(\rv).
       \label{EQGintGeneralDefinition}
\end{align}

In our planar geometrical setup we consider single-phase systems under
the influence of external force fields $\fv_\rmext^{(\alpha)}(\rv)=
f_\rmext^{(\alpha)}(z)\ev_z$, where $\ev_z$ is the unit vector in the
$z$-direction.
In the chosen geometry the particles move with mean velocity
$\vel_\alpha(z) = v_\alpha(z)\ev_z$. The counterforce density acts
parallel to the motion, $\Gv_\rmint(\rv)=G_\rmint(z)\ev_z$, and it
represents the emerging drag that the particles of the two oppositely
moving species exert onto each other.
The general power functional dependence simplifies to
$G_\rmint(z;[v_\Delta, \rho_1,\rho_2])$, where the velocity difference
$v_\Delta(z) = v_2(z)-v_1(z)$ is a measure of the degree of the
counterflow in the system.
We first consider cases with force amplitudes
$f_\rmext^\alphasup(z)={\rm const}$, such that also the partial
density profiles $\rho_\alpha(z) =\rm const$.  For such homogeneous
systems, as are illustrated in Fig.~\ref{FIGsnapshot}, it follows that
also $G_\rmint(z)=\rm const$.  As all pertinent quantities are
spatially constant in the considered homogeneous case, the functional
dependence simplifies to a {\it parametric} dependence:
$G_\rmint(z;[v_\Delta, \rho_1,\rho_2]) = {\cal G}(v_\Delta, \rho_1,
\rho_2)$

\begin{figure}[!t]
  \vspace{1mm}
  \includegraphics[page=1,width=.99\columnwidth]{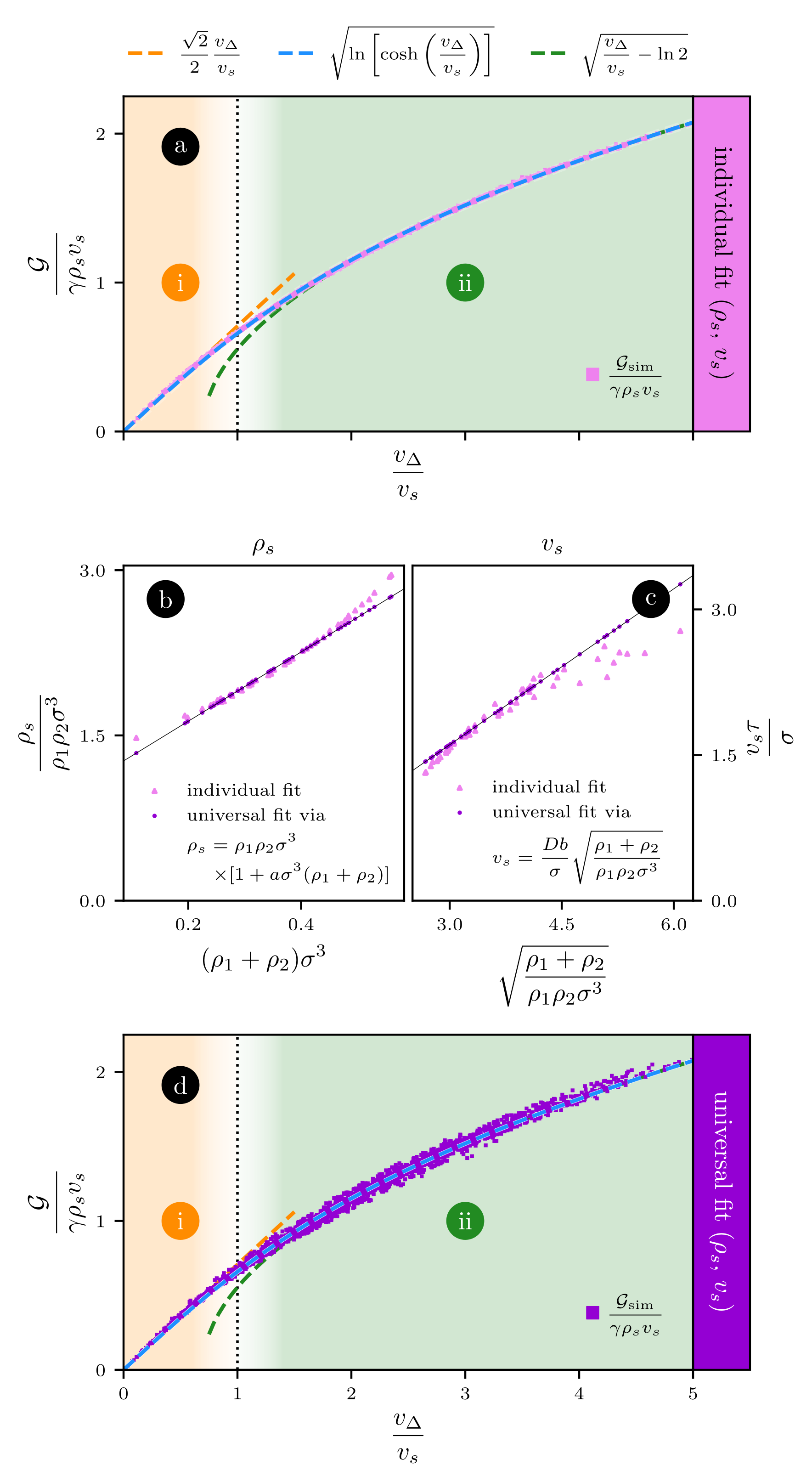}
  \caption{Nonequilibrium scaling of drag forces.  a) Drag force ratio
    $\calG/(\gamma\rho_s v_s)$ shown as a function of the counterdrive
    velocity ratio $v_\Delta/v_s$. Simulation results are obtained
    over wide ranges of partial densities and driving strengths, see
    Fig.~\ref{FIGsnapshot}. The `individual fit' results (pink data
    points) collapse onto the general scaling function
    \eqref{EQscaling} (dashed blue curve). Indicated is the asymptotic
    behaviour for i)~near-equilibrium (dashed orange) and
    ii)~far-nonequilibrium (dashed green), see \eqr{EQasymptotic}.
    b) Scaled effective density $\rho_s/(\rho_1\rho_2\sigma^3)$ as a
    function of the scaled total density $(\rho_1+\rho_2)\sigma^3$,
    obtained from the individual fits (pink symbols) and for the
    `universal fit' (thin black line), where the data points (violet
    symbols) agree by construction with \eqr{EQrhos}.
    c) Same as b) but for the scaled crossover velocity $v_s
    \tau/\sigma$ as a function of
    $\sqrt{(\rho_1+\rho_2)/(\rho_1\rho_2\sigma^3)}$ and compared
    against \eqr{EQvels}.
    d) Same as a) but obtained from the universal fit. Despite some
    scatter the correct scaling is retained.}
\label{FIGscaling}
\end{figure}

To make progress, we propose the following ansatz for the scaling
behaviour of the uniform drag force density:
\begin{align}
  {\cal G}(v_\Delta, \rho_1, \rho_2) &= 
  \gamma  \rho_s v_s 
  \sqrt{\ln\cosh\left(v_\Delta/v_s\right)} \sgn v_\Delta,
  \label{EQscaling}
\end{align}
and we lay out a justification based on asymptotics below.
In the ansatz \eqref{EQscaling} the velocity scale
$v_s(\rho_1,\rho_2)$ and the effective density $\rho_s(\rho_1,\rho_2)$
are as-yet unknown functions, which are amenable to data-driven modelling, as demonstrated below. These encapsulate the dependence on the
partial densities $\rho_1$ and $\rho_2$ and the functions play
decisive roles in the behaviour of the scaling relationship
\eqref{EQscaling}.  The velocity scale $v_s$ determines where the
regimes of low- and high-drive behaviour cross over. The effective
particle density $\rho_s$ governs the overall amplitude of the drag
force density~\eqref{EQscaling}.  The spatial symmetry is odd, such
that reversing the flow direction reverses the direction of the drag
force density, ${\cal G}(-v_\Delta, \rho_1, \rho_2)=-{\cal
  G}(v_\Delta, \rho_1, \rho_2)$.

The functional form of Eq.~\eqref{EQscaling} is deliberately chosen to cross over from linear to sublinear scaling for increasing counterdrive velocity, as is consistent with the simulation evidence presented below.
In particular, the scaling function \eqref{EQscaling} allows one to match
simultaneously the behaviour in both the near-equilibrium and
far-nonequilibrium regimes, as we demonstrate in the following. We
first address the near-equilibrium behaviour and thus Taylor expand
\eqr{EQscaling} in the small parameter $x=v_\Delta/v_s$.  To lowest
order this gives $\sqrt{\ln\cosh
  x}=\sqrt{\ln(1+x^2/2+O(x^4))}=\sqrt{x^2/2+O(x^4)}=|x|/\sqrt{2}+O(x^3)$.
We hence obtain the simple linear-response result
$\calG(v_\Delta;\rho_1,\rho_2) \sim \gamma \rho_s v_\Delta / \sqrt{2}$, where the
tilde indicates equality in the scaling regime. We recall that
$\rho_s(\rho_1,\rho_2)$ is independent of $v_\Delta$. Thus the
near-equilibrium behaviour is independent of the crossover velocity
$v_s$ and bears no traces of its restricted validity towards the
nonlinear regime.

We address the far-nonequilibrum behaviour by performing an asymptotic
expansion of \eqr{EQscaling} for $x\to\pm\infty$, as follows from
approximating the hyperbolic cosine: $\sqrt{\ln \cosh x}\sim
\sqrt{\ln(\e^{|x|}/2)}=\sqrt{|x|-\ln 2}$.  The resulting
far-nonequilibrium form of \eqr{EQscaling} is ${\cal G} \sim \gamma
\rho_s v_s \sqrt{|v_\Delta|/v_s -\ln 2}\sgn v_\Delta$.  For high
counterflow, $v_\Delta/v_s\to\pm \infty$, the dominant term is $\gamma
\rho_s v_s \sqrt{|v_\Delta|/v_s}\sgn v_\Delta$. Notably the dependence
on the crossover velocity $v_s$ is retained in this limit. This
implies that from studying the far-nonequilibrium behaviour only, one
can infer its range of validity towards the near-equilibrium regime.
Summarizing the above results for the low- and high-flow regimes, we
have
\begin{align}
  \frac{{\cal G}(v_\Delta, \rho_1, \rho_2)}{\gamma\rho_s v_s} &\sim
  \begin{cases}
    v_\Delta/(\sqrt{2} v_s) & |v_\Delta| \ll v_s\\
    \sqrt{|v_\Delta|/v_s-\ln 2}\sgn v_\Delta
    & v_s \ll |v_\Delta|,
  \end{cases}
  \label{EQasymptotic}
\end{align}
where we recall the tilde to indicate equality in the respective
scaling regime.

We illustrate the different regimes by plotting the ratio ${\cal
  G}(v_\Delta,\rho_1,\rho_2)/(\gamma\rho_s v_s)$ in full form
\eqref{EQscaling}, together with its two asymptotically leading
expansions~\eqref{EQasymptotic}, as a function of the scaled velocity
difference~$v_\Delta/v_s$ in Fig.~\ref{FIGscaling}a.  The results from
our adaptive Brownian dynamics simulations are obtained under
spatially constant driving conditions over considerable ranges of
values of the partial densities $\rho_1$ and $\rho_2$, which vary
independently from $0.03\sigma^{-3}$ to $0.30\sigma^{-3}$, and with
different driving strengths $f_{\rmext}^\alphasup$ up to
$6\epsilon/\sigma$, as depicted in Fig.~\ref{FIGsnapshot}.

To connect the theory quantitatively with the simulation results we
first build subsets of all simulated systems that share the same pair
of bulk density values $\rho_1$ and~$\rho_2$. In each subset we treat
the two scaling parameters $\rho_s$ and $v_s$ as `individual fit'
parameters to describe the counterflow behaviour in the subset. The
resulting agreement of the simulation data with the scaling
form~\eqref{EQscaling} is near-perfect, see
Fig.~\ref{FIGscaling}a. Similarly good agreement of the theoretical
model and the simulation results, with some slight increase in scatter
that is apparent in Fig.~\ref{FIGscaling}d, is obtained from the
following alternative `universal fit'.

The universal fit is obtained by observing that the dependence of the
scaling parameters $\rho_s$ and $v_s$ on the partial densities
$\rho_1$ and $\rho_2$ turns out to be relatively simple. We show in
Fig.~\ref{FIGscaling}b the behaviour of the scaled crossover density
$\rho_s/(\rho_1\rho_2\sigma^3)$ as a function of the scaled total
density $(\rho_1+\rho_2)\sigma^3$. Correspondingly we show in
Fig.~\ref{FIGscaling}c the scaled crossover velocity $v_s\tau/\sigma$
as a function of the density combination
$\sqrt{(\rho_1+\rho_2)/(\rho_1\rho_2\sigma^3)}$. In both cases linear
behaviour is found,
which can be {\it quantitatively} represented by the following simple
`universal fit' forms:
\begin{align}
  \rho_s(\rho_1,\rho_2) &= \rho_1\rho_2 
      [1+a(\rho_1+\rho_2)\sigma^3] \sigma^3,
      \label{EQrhos}\\
  v_s(\rho_1,\rho_2) &= 
  \frac{Db}{\sigma} 
  \sqrt{\frac{\rho_1+\rho_2}{\rho_1\rho_2\sigma^3}}.
  \label{EQvels}
\end{align}
Here $a\sigma^3$ constitutes an interaction volume that connects the
partial densities with the effective density $\rho_s$ and $b/\sigma$
is an inverse interaction length scale that relates the free diffusion
constant (we recall $D=k_BT/\gamma$) to the crossover velocity~$v_s$.
We obtain the parameter values $a=3.14$ and $b=1.07$ from a
least-squares fit across our entire data set.  The thin black lines in
Fig.~\ref{FIGscaling}b,~c indicate these choices in
Eqs.~\eqref{EQrhos} and \eqref{EQvels}. Despite some outliers and the
agreement worsening slightly for very dissimilar partial densities,
the full behaviour of the simulation results is captured
quantitatively by the analytical forms~\eqref{EQrhos}
and~\eqref{EQvels}.

We next compare the results of the analytical counterforce model, as
given by the scaling form \eqref{EQscaling} of ${\cal
  G}(v_\Delta,\rho_1,\rho_2)$ and the parameter dependencies
\eqref{EQrhos} and~\eqref{EQvels},
against the results of all our simulations. Although the simple
expressions \eqref{EQrhos} and \eqref{EQvels} introduce some
deviations from the simulations results, the overall level of
agreement is very satisfactory, see Fig.~\ref{FIGscaling}d. We recall
the large parameter space that is explored by the simulations, as
illustrated in Fig.~\ref{FIGsnapshot}. The overall `universal' model
\eqref{EQscaling}, \eqref{EQrhos} and \eqref{EQvels} thus contains
merely two empirical parameters $a$ and $b$ and it captures
quantitatively the {\it entirety} of our simulation results for
homogeneous conditions near to and far from equilibrium.

\begin{figure}[t!]
  \hspace{1mm}
  \includegraphics[page=1,width=\columnwidth]{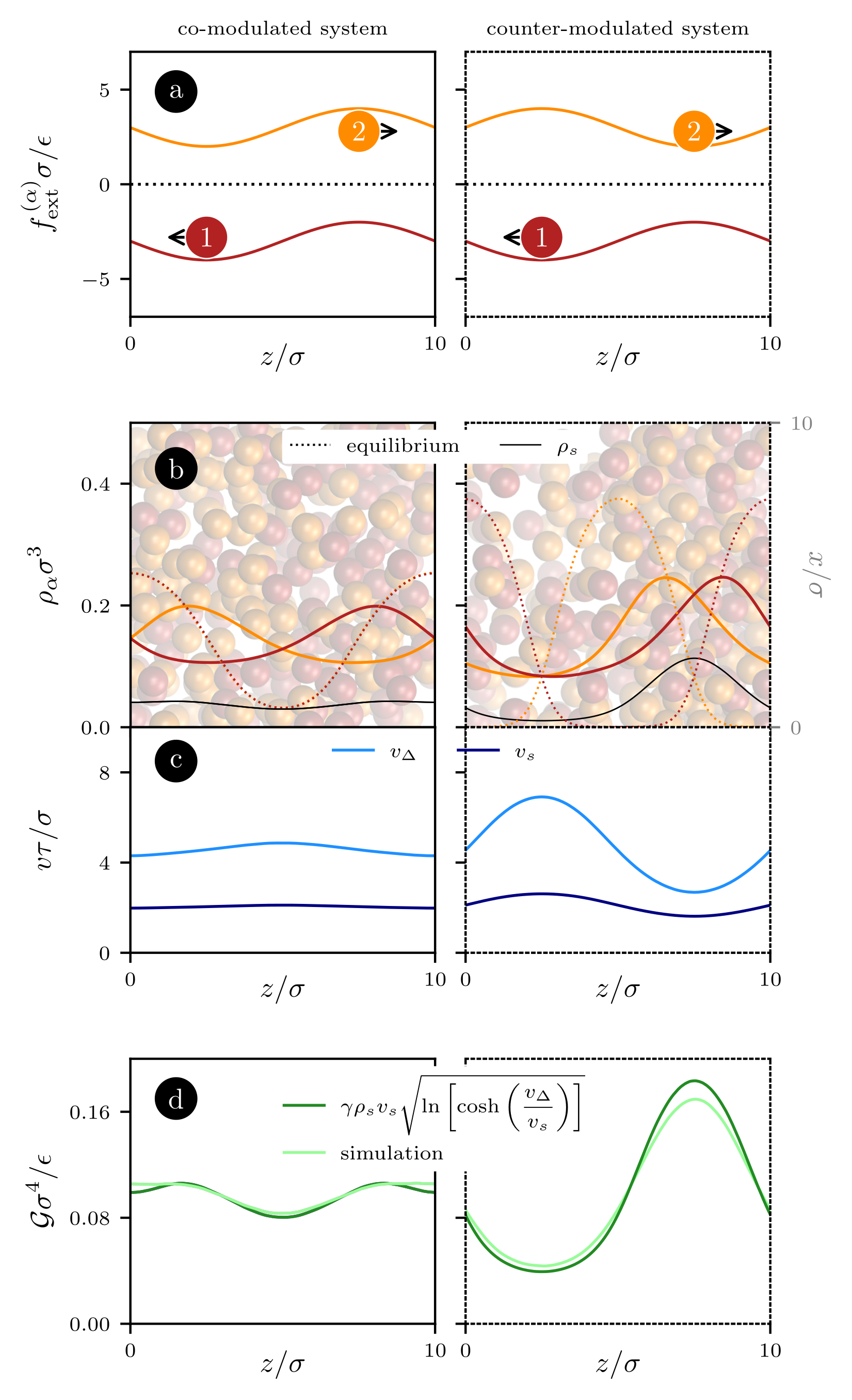}
  \caption{Inhomogeneous systems with co-modulated (left column) and
    counter-modulated driving (right column) along the
    $z$-direction. a)~The inhomogeneous external force fields
    $f_\rmext^\alphasup(z) \sigma/\epsilon$ for species $\alpha=1$
    (dark red) and 2 (light orange). b)~The resulting species-resolved
    nonequilibrium density profiles $\rho_1(z)\sigma^3$ and
    $\rho_2(z)\sigma^3$ together with the corresponding equilibrium
    density profiles (dashed lines) and the local scaling density
    $\rho_s(z)\sigma^3$ (black line). The background shows
    representative simulation snapshots (with horizontal
    $z$-direction) to illustrate the spatial concentration and density
    variations. c)~The local velocity difference profile $v_\Delta(z)
    \tau/\sigma$ (light blue line), as obtained from simulations,
    together with the local scaling velocity $v_s(z) \tau/\sigma$
    (dark blue line). d)~The relative motion of the two species is
    well described by the drag force density
    $\calG(z)\sigma^4/\epsilon$ obtained from the local power
    functional~\eqref{EQinhomogeneous} (dark green lines), as
    confirmed by comparison to simulation results (light green lines).
\label{FIGinhomogeneous}}
\end{figure}

We use the validity of this analytical model for conditions of
spatially homogeneous drive as an outset to develop a theory that
captures long-ranged spatial variation in terms of a local power
functional approximation. We address two representative model
situations.  The first, `co-modulated' case is colour-blind such that
the constant counterdrive $\pm f_0$ is supplemented by a modulated
force field that is identical for both species.  The external force
fields are $f_\rmext^\alphasup(z)= \pm f_0 - f_1 \sin(2\pi z/L)$,
where $f_0$ is the offset, $f_1$ is an amplitude parameter, and $\pm$
refers to species $\alpha= 1,2$.  In the second, `counter-modulated'
case the entire external force acts in opposite directions for both
species, which is akin to a symmetric electrolyte in an electric
field.  The external force fields are $f_\rmext^\alphasup(z)= \pm f_0
\pm f_1 \sin(2\pi z/L)$.  In both cases the two species are driven
against each other by the constant force~$\pm f_0$. We show
representative results for $f_0=3 \epsilon/\sigma$ and $f_1=1
\epsilon/\sigma$.

In the co-modulated system (Fig.~\ref{FIGinhomogeneous}, left column)
the drag force exhibits weak spatial variation that displays maxima in
the regions of maximal density difference between the two species.
The counter-modulated system (Fig.~\ref{FIGinhomogeneous}, right
column) displays much stronger modulation of the density contrast, as
one might expect from the setup. The counterflow acts to move the two
density maxima much closer together than they are in equilibrium
($f_0=0$). The spatial region around the two partial density maxima
has a strong peak in the local drag force density. The shape of the
peak is spatially more localized than the shallow minimum
that occurs in the region of relative density depletion.

We rationalize these simulation results on the basis of a theoretical
model for {\it inhomogeneous} counterflow based on the above scaling
forms~\eqref{EQscaling}, \eqref{EQrhos}, and \eqref{EQvels} that were
identified for homogeneous drive.  We use a local power functional
approximation, where we let the scaling parameters depend on the
partial density profiles via localized versions of Eqs.~\eqref{EQrhos}
and \eqref{EQvels}, such that the local density scale is
$\rho_s(z)=\rho_s(\rho_1(z),\rho_2(z))$, and the local crossover
velocity is $v_s(z)=v_s(\rho_1(z),\rho_2(z))$.  This allows one to
express a local approximation of the universal scaling function
\eqref{EQscaling} in the form:
\begin{align}
  \calG(z) 
  &={\cal G}\big(v_\Delta(z), \rho_1(z), \rho_2(z)\big).
  \label{EQinhomogeneous}
\end{align}
When compared against simulation results, the local power functional
approximation \eqref{EQinhomogeneous} performs very satisfactorily, as
is shown for the two representative inhomogeneous systems in
Fig.~\ref{FIGinhomogeneous}.  The scaling model captures correctly the
physics of inhomogeneous conditions via the power functional
dependence of the mean interparticle force field on the counterflow
velocity. Any mere dynamical density functional dependence misses the
effect by construction, which is due to the omission of superadiabatic (i.e.\ genuine out-of-equilibrium) contributions \cite{schmidt2013pft,schmidt2022rmp} to the interparticle force density. This approximation is uncontrolled in general, and in the present application, the inclusion of superadiabatic forces turns out to be decisive for modelling accurately the emerging counterdriven flow, as is facilitated by the power functional approach.

In conclusion we have presented simulation-based evidence for the
nonequilibrium scaling of drag forces in driven mixtures. For the
considered steady states our results confirm the existence of the
power functional map \cite{schmidt2022rmp, schmidt2013pft,
  brader2015dtpl} from the partial density and velocity profiles to
the localized one-body force densities via the explicit functional
independence on the external forces. Based on a systematic force
decomposition~\cite{geigenfeind2020laning}, our work goes beyond
previous efforts in modelling the functional dependencies. Here we
have identified both near-equilibrium and far-nonequilibrium regimes
and elucidated quantitatively the intervening crossover behaviour.

Future work could address the possible generality of the proposed
scenario beyond counterflow and consider methods for first-principles
derivations as well as nonequilibrium functional machine learning
\cite{delasheras2023perspective, zimmermann2024ml}, where binary
mixtures have been addressed successfully in equilibrium
\cite{robitschko2025mixShort, zhou2026azeoptropic, bui2024neuralrpm,
  bui2025dielectrocapillarity}.
We emphasize that we find it highly relevant to identify microscopic mechanisms
for the observed far-nonequilibrium weakening of the drag force scaling, which
we determine to be of square-root type. While our current evidence for this
behaviour is empirical and based on the analysis of simulation data, we
find this to be a well-controlled yet highly nontrivial and potentially
prototypical effect that warrants the application and development of advanced
nonequilibrium statistical mechanical methodology.
It remains to be scrutinized whether the identified scaling continues to apply for different types of dynamics, e.g.\ when incorporating hydrodynamic interactions into the underlying stochastic equations of motion, or when considering long-ranged (e.g.\ electrostatic) interparticle interactions, as is relevant to address transport in electrolytes.
In particular for inhomogeneous systems, the resulting nonequilibrium response in such extended scenarios may be subtle and needs to be addressed in future work.
It would further be compelling to calculate linear response properties from first principles, e.g.\ based on the Green-Kubo response function formalism.

It is also interesting to investigate
the significance of our findings in light of current research
dedicated to the physics of responsive colloidal particles that can
adapt their interparticle forces through a variety of different
influences and across varying physical
situations~\cite{monchojorda2020, bley2021, bley2022, baul2021,
  lopezmolina2024, monchojorda2023}. As our methodology rests firmly
on the one-body force balance equation and functional concepts, it
could be highly rewarding to relate the present findings to
hyperforces \cite{robitschko2024any,matthes2024mix} in the context of
the recent gauge invariance of statistical mechanics
\cite{hermann2021noether, mueller2024gauge, mueller2024dynamic,
  mueller2024whygauge, maruyama2026, nguyen2026, phamvan2026symmetry}.

{\it Data availability}---%
The data that support the findings of this study were generated by numerical simulations.
The source code and parameters used to generate the simulations is publicly available \cite{MBD}.

\begin{acknowledgments}
  We thank Silas Robitschko and Pia Fleischmann for useful discussions.
  Calculations were performed using the emil-cluster of the Bayreuth Center for High Performance Computing funded by the DFG (Deutsche Forschungsgemeinschaft) under project no.~422127126.
  This work is supported by the DFG (Deutsche Forschungsgemeinschaft) under project no.~551294732.
\end{acknowledgments}

\bibliography{noe}

\end{document}